\documentclass[preprint,prb]{revtex4}
\usepackage{amsmath}

\newcommand{\gradv}{\boldsymbol{\nabla}}
\def\v#1{{\bf#1}}

\begin{document}

\title{The exact relation between the displacement current and the conduction current: Comment on a paper by Griffiths and Heald}
\author{Jos\'e A. Heras}
\email{herasgomez@gmail.com}
\affiliation{Departamento de Ciencias B\'asicas, Universidad Aut\'onoma Metropolitana, Unidad Azcapotzalco, Av. San Pablo No. 180, Col. Reynosa, 02200, M\'exico D. F. M\'exico and Departamento de F\'isica y Matem\'aticas, Universidad Iberoamericana, Prolongaci\'on Paseo de la Reforma 880, M\'exico D. F. 01210, M\'exico}

\begin{abstract}
I introduce the exact relation between the displacement current $\epsilon_0 \partial \v E/\partial t$ and the ordinary current $\v J$. I show that $\epsilon_0 \partial \v E/\partial t$ contains a local term determined by the present values of $\v J$ plus a non-local term determined by the retarded values of $\v J$. The non-local term implements quantitatively the suggestion made by Griffiths and Heald that
the displacement current at a point is a surrogate for ordinary currents at other locations.
\end{abstract}
\maketitle

Griffiths and Heald have posed and answered the following question (their ``recurring question (3)"):\cite{1} ``What relationship exists between the two sources in the Ampere-Maxwell law:
\begin{equation}
\gradv\times \v B=\mu_0\v J +\mu_0\epsilon_0\frac{\partial \v E}{\partial t},
\end{equation}
for instance, can one be derived from the other?" Their answer was the following:\cite{1} ``None. The two currents are independent and both are needed, in general. $\v J$ represents the ordinary current at a point where the curl [of $\v B]$ is evaluated, while the ``displacement current" $\epsilon_0 \partial \v E/\partial t$ is a surrogate for ordinary currents at other locations." The assumed lack of a specific relation between $\epsilon_0 \partial \v E/\partial t$ and $\v J$ led to the claim that the Biot-Savart law extended to include the displacement current:
\begin{equation}
\v B=\frac{\mu_0}{4\pi}\int d^3x'\frac{(\v J+\epsilon_0\partial \v E/\partial t)\times{\hat{\v R}}}{R^2},
\end{equation}
``\ldots\ is not useful because it is self-referential: To calculate $\v B$ at a point, one must know $\v E$ $everywhere$ -- but you can't determine $\v E$ unless you already know $\v B$ (everywhere), because of Faraday induction $[\gradv\times\v E=-\partial \v B/\partial t].$" In Eq.~(2) the integration is over all space and $\hat{\v R}={\v R}/R =(\v x-\v x')/|\v x-\v x'|$.

In this comment I derive the exact relation between the displacement current $\epsilon_0\partial \v E/\partial t$ and the current $\v J$. The existence of this relation allows us to answer in the affirmative the question posed and answered negatively by Griffiths and Heald.\cite{1}

Let me first introduce this relation which we will derive in the following:
\begin{equation}
\epsilon_0\frac{\partial \v E}{\partial t}\! =\! 
-\frac{\v J}{3}+ \frac{1}{4\pi}\!\int\! d^3x'\bigg(\frac{3\hat{\v R}(\hat{\v R} \cdot [\v J])\!  -\!  [\v J]}{R^3}
+ \frac{3\hat{\v R}(\hat{\v R} \cdot [\partial \v J/\partial t])\!  -\!  [\partial \v J/\partial t]}{R^2 c} + \frac{\hat{\v R}\!  \times\!  (\hat{\v R} \! \times \! [\partial^2 \v J/\partial t^2]}{Rc^2} \bigg),
\end{equation}
where the square brackets $[\;]$ indicate that the enclosed quantity is to be evaluated at the retarded time $t'=t-R/c$. Equation~(3) provides the exact relation between the currents $\epsilon_0\partial \v E/\partial t$ and $\v J$ and implies that the two currents are not independent. The first term represents a local part which exists in the region where $\v J$ is defined and depends on the present values of $\v J$. The second term is a non-local part which extends over all space and depends on the retarded values of $\v J$. In short, the local contribution to $\epsilon_0\partial \v E/\partial t$ is instantaneously produced by $\v J$, and the non-local contribution to $\epsilon_0 \partial \v E/\partial t$ is causally produced by $\v J$. It follows that the displacement current is produced by the present and retarded values of the ordinary current $\v J$. The non-local part of $\epsilon_0 \partial \v E/\partial t$ shows explicitly how the displacement current at a point represents the contribution from ordinary currents elsewhere. Equation (3) implements quantitatively the suggestion made by Griffiths and Heald that:\cite{1} ``the displacement current at a point is a surrogate for ordinary currents at other locations." These authors also claimed that Eq.~(2) is self-referential. However, the substitution of Eq.~(3) into Eq.~(2) removes the self-referential aspect of Eq.~(2). To calculate $\v B$ by means of Eq.~(2) we really need to know $\epsilon_0\partial\v E/\partial t$ and not necessarily $\v E$, and we can calculate $\epsilon_0\partial\v E/\partial t$ from $\v J$ using Eq.~(3) with no explicit reference to $\v B$.\cite{2}

To obtain Eq.~(3) we first consider the generalized Biot-Savart law:\cite{3}
\begin{equation}
 \v B= \frac{\mu_0}{4\pi}\!\int\! d^3x' \bigg(\frac{[\v J]\times\hat{\v R}}{R^2 }+\frac{[\partial \v J/\partial t]\times\hat{\v R}}{R c}\bigg).
\end{equation}
We next take the curl of Eq.~(4)
\begin{equation}
\nabla\times\v B = \frac{\mu_0}{4\pi}\!\int\! d^3x' \bigg(\gradv\times \bigg\{\frac{[\v J]\times\hat{\v R}}{R^2 }\bigg\}+\gradv\times\bigg\{\frac{[\partial \v J/\partial t]\times\hat{\v R}}{R c}\bigg\}\bigg).
\end{equation}
The following identities are demonstrated in Appendix A:
\begin{align}
\gradv\times \bigg\{\frac{[\v J]\times\hat{\v R}}{R^2 }\bigg\} & = \frac{3\hat{\v R}(\hat{\v R}\cdot[\v J]) -[\v J]}{R^3}+\frac{\hat{\v R}(\hat{\v R}\cdot[\partial \v J/\partial t])-[\partial \v J/\partial t]}{R^2 c}+\frac{8\pi}{3}\v [\v J]\delta(\v x-\v x'),\\
\gradv\times\bigg\{\!\frac{[\partial \v J/\partial t]\times\hat{\v R}}{R c}\bigg\} & =\frac{2\hat{\v R}(\hat{\v R}\cdot[\partial \v J/\partial t])}{R^2 c}+\frac{\hat{\v R} \times (\hat{\v R} \times [\partial^2 \v J/\partial t^2]}{Rc^2},
\end{align}
where $\delta(\v x-\v x')$ is the Dirac-delta function.
If we insert Eq.~(6) [with its last term written as $4\pi\v [\v J]\delta(\v x - \v x') - (4\pi/3)\v [\v J]\delta(\v x- \v x')$] and Eq.~(7) into Eq.~(5), and integrate the Dirac-delta terms, we obtain
\begin{align}
\nabla\times\v B & = \mu_0\v J+\mu_0\Bigg\{-\frac{\v J}{3}+ \frac{1}{4\pi}\!\int\! d^3x'\bigg(\frac{3\hat{\v R}(\hat{\v R}\cdot [\v J]) - [\v J]}{R^3}
+ \frac{3\hat{\v R}(\hat{\v R} \cdot [\partial \v J/\partial t]) - [\partial \v J/\partial t]}{R^2 c}
\nonumber\\
&{} \qquad + \frac{\hat{\v R} \times (\hat{\v R} \times [\partial^2 \v J/\partial t^2]}{Rc^2} \bigg)\!\Bigg\}.
\end{align}
The term within the parentheses $\{\;\}$ is identified as the term $\epsilon_0\partial\v E/\partial t$ in Eq.~(1). From this identification we obtain Eq.~(3).

Equation (3) can be applied to the polarization current $\partial\v P/\partial t$, where $\v P$ is a given polarization density. The replacement $\v J\to \!\partial\v P/\partial t$ in Eq.~(3) and the fact that $\partial/\partial t$ can be taken outside the retardation parentheses and even outside the integral leads to
\begin{align}
&\epsilon_0\frac{\partial \v E}{\partial t} = \frac{\partial }{\partial t}\Bigg\{
-\frac{\v P}{3} + \frac{1}{4\pi}\!\int\! d^3x'\bigg(\frac{3\hat{\v R}(\hat{\v R}\!\cdot\![\v P]) - [\v P]}{R^3}
\nonumber\\&\qquad\qquad+ \frac{3\hat{\v R}(\hat{\v R} \cdot [\partial \v P/\partial t])\!-\![\partial \v P/\partial t]}{R^2 c}
\!+\!\frac{\hat{\v R} \times (\hat{\v R} \times [\partial^2 \v P/\partial t^2]}{Rc^2}\bigg) \Bigg\},
\end{align}
which implies that the electric field produced by the polarization density is\cite{4}
\begin{equation}
\v E \!=\! -\frac{\v P}{3\epsilon_0} + \frac{1}{4\pi\epsilon_0}\!\int\! d^3x'\!\bigg(\frac{3\hat{\v R}(\hat{\v R}\!  \cdot\!  [\v P])\! -\! [\v P]}{R^3}
 + \frac{3\hat{\v R}(\hat{\v R} \! \cdot \! [\partial \v P/\partial t])\! -\! [\partial \v P/\partial t]}{R^2 c}
 + \frac{\hat{\v R} \! \times \! (\hat{\v R}\!  \times\!  [\partial^2 \v P/\partial t^2]}{Rc^2} \bigg).
\end{equation}

We can apply Eq.~(10) to a point electric dipole possessing a time varying dipole moment $\v p(t)$ located at the point $\v x_0$. The polarization density for such a dipole is given by $\v P(\v x,t)= \v p(t)\delta\{\v x-\v x_0\}$. After substituting this polarization density into Eq.~(10) and integrating over all space, we find the Hertzian electric field:
\begin{equation}
\v E = -\frac{\v p(t)}{3\epsilon_0}\delta\{\v x - \v x_0\} + \frac{1}{4\pi\epsilon_0} \bigg(\frac{3\hat{\v R}(\hat{\v R} \cdot [\v p]) - [\v p]}{R^3}
+ \frac{3\hat{\v R}(\hat{\v R} \cdot [\dot{\v p}]) - [\dot{\v p}]}{R^2 c}
 + \frac{\hat{\v R} \times (\hat{\v R} \times [\ddot{\v p}]}{Rc^2}\!\bigg),
\end{equation}
where $\hat{\v R}={\v R}/R =(\v x-\v x_0)/|\v x-\v x_0|$, $\dot{\v p}=\partial \v p/\partial t$,
and the square brackets $[\;]$ now indicate that the enclosed quantity is to be evaluated at the retarded time $t'=t-|\v x-\v x_0|/c$. The electric field in Eq.~(11) is usually presented without the delta function term.

The origin of the local term $\v J/3$ in Eq.~(3) might seem intriguing. However, if we use the Panofsky and Phillips formula for the retarded electric field:\cite{5}
\begin{equation}
\v E=\frac{1}{4\pi\epsilon_0}\!\int\! d^3x'\!\bigg(\frac{[\rho]\hat{\v R}}{R^2}+\frac{2\hat{\v R}(\hat{\v R}\cdot[\v J])-[\v J]}{R^2c}
+\frac{\hat{\v R}\times(\hat{\v R}\times[\partial \v J/\partial t]}{Rc^2} \bigg),
\end{equation}
we can show the origin of the $\v J/3$ term. Equation (12) is an equivalent expression of the generalized Coulomb law.\cite{3} The time derivative of Eq.~(12) yields
\begin{equation}
\epsilon_0\frac{\partial \v E}{\partial t}=\frac{1}{4\pi}\!\int\! d^3x'\frac{[\partial\rho/\partial t]\hat{\v R}}{R^2}
+\frac{1}{4\pi}\!\int\! d^3x'\!\bigg(\frac{2\hat{\v R}(\hat{\v R}\cdot[\partial\v J/\partial t])-[\partial \v J/\partial t]}{R^2c}
+\frac{\hat{\v R}\times(\hat{\v R}\times[\partial^2 \v J/\partial t^2]}{Rc^2} \!\bigg).
\end{equation}
We use the continuity equation and write the first integral as (Appendix A)
\begin{equation}
\int\! d^3x'\frac{[\partial\rho/\partial t]\hat{\v R}}{R^2}=-
\frac{4\pi}{3} \v J +\!\int\! d^3x'\bigg(\frac{3\hat{\v R}(\hat{\v R}\cdot[\v J]) - [\v J]}{R^3} + \frac{\hat{\v R}(\hat{\v R} \cdot [\partial\v J/\partial t])}{R^2c}\bigg).
\end{equation}
The insertion of Eq.~(14) into Eq.~(13) gives Eq.~(3) again.

The existence of the $\v J/3$ term in Eq.~(3) can be traced to the first term in Eq.~(14). Thus, the origin of the $\v J/3$ term is attributable to time variations of the charge density and ultimately to the continuity equation (charge conservation).
I invite readers to decide which of the two derivations of Eq.~(3) presented here they find
more useful.

\appendix
\section{Derivation of Eqs.~(6), (7), and (14)}
Consider the expansion
\begin{subequations}
\begin{align}
\gradv \times \bigg\{ \frac{[\v J] \times \hat{\v R}}{R^2 }\bigg\} &= - \gradv\times \bigg\{[\v J]\times\gradv\bigg(\frac{1}{R}\bigg)\bigg\}\\
&= - [\v J]\nabla^2\bigg(\frac{1}{R}\bigg) + \gradv\bigg(\frac{1}{R}\bigg)\gradv\cdot[\v J] - \bigg(\gradv\bigg(\frac{1}{R}\bigg)\cdot\gradv\bigg)[\v J] + \bigg([\v J]\cdot\gradv\bigg)\gradv\bigg(\frac{1}{R}\!\bigg).
\end{align}
\end{subequations}
It can be shown that\cite{6}
\begin{subequations}
\begin{align}
\nabla^2\bigg(\!\frac{1}{R}\!\bigg)&=-4\pi\delta(\v x-\v x'),\\
\gradv\cdot[\v J]&=-\frac{\hat{\v R}\cdot[\partial \v J/\partial t]}{c},\\
\bigg(\gradv\bigg(\frac{1}{R}\bigg)\cdot\gradv\bigg)[\v J]&=\frac{[\partial \v J/\partial t]}{R^2 c},\\
\bigg([\v J]\cdot\gradv\bigg)\gradv\bigg(\frac{1}{R}\bigg)&=\frac{3\hat{\v R}(\hat{\v R}\cdot[\v J]) - [\v J]}{R^3}-\frac{4\pi}{3}[\v J]\delta(\v x-\v x').
\end{align}
\end{subequations}
To obtain Eq.~(A2d) we used the identity:\cite{7}
\begin{equation}
\frac{\partial}{\partial x_i\partial x_j}\bigg(\frac{1}{R}\bigg)=\frac{3\hat{R_i}\hat{R_j}-\delta_{ij}}{R^3} -\frac{4\pi}{3}\delta_{ij}\delta(\v x-\v x'),
\end{equation}
where $\hat{R_i}=R_i/R=(x_i-x'_i)/|x_i-x'_i|$ and $\delta_{ij}$ is the Kronecker delta symbol. The substitution of Eq.~(A2) into Eq.~(A1) leads directly to Eq.~(6).

Consider the expansion
\begin{eqnarray}
\gradv \times \bigg\{\frac{[\partial \v J/\partial t] \times \hat{\v R}}{Rc}\bigg\} &=& [\partial \v J/\partial t]\gradv \cdot \bigg(\frac{\hat{\v R}}{Rc}\bigg) - \bigg(\frac{\hat{\v R}}{Rc}\bigg)\gradv \cdot [\partial \v J/\partial t] \notag \\
&&{}+ \bigg(\frac{\hat{\v R}}{Rc}\cdot\gradv\bigg)[\partial \v J/\partial t] - \bigg( [\partial \v J/\partial t] \cdot \gradv\bigg) \frac{\hat{\v R}}{Rc}.
\end{eqnarray}
If we substitute the results
\begin{subequations}
\begin{align}
\gradv \cdot\bigg(\frac{\hat{\v R}}{Rc}\bigg)&=\frac{1}{R^2c},\\
\gradv\cdot[\partial \v J/\partial t]&=-\frac{\hat{\v R}\cdot[\partial^2 \v J/\partial t^2]}{c},\\
\bigg(\frac{\hat{\v R}}{Rc}\cdot\gradv\bigg)[\partial \v J/\partial t]&=-\frac{[\partial^2 \v J/\partial t^2]}{R c^2},\\
\bigg([\partial \v J/\partial t]\cdot\gradv\bigg) \frac{\hat{\v R}}{Rc}&=-\frac{2\hat{\v R}(\hat{\v R}\cdot[\partial \v J/\partial t]) - [\partial \v J/\partial t]}{R^2 c},
\end{align}
\end{subequations}
into Eq.~(A4), we obtain Eq.~(7).

The proof of Eq.~(14) goes as follows: If the continuity equation and Eq.~(A2b) are substituted into the identity $[\gradv'\cdot\v J]= \gradv'\cdot[\v J] +\gradv\cdot[\v J]$ and the resulting equation multiplied by $\hat{\v R}/R^2$, we obtain
\begin{equation}
\frac{[\partial\rho/\partial t]\hat{\v R}}{R^2} = - \frac{(\gradv'\cdot[\v J])\hat{\v R}}{R^2} + \frac{\hat{\v R}(\hat{\v R}\cdot[\partial \v J/\partial t])}{R^2c},
\end{equation}
and hence
\begin{equation}
\int\!d^3x'\frac{[\partial\rho/\partial t]\hat{\v R}}{R^2} =
 -\!
\int\! d^3x'\frac{(\gradv'\cdot[\v J])\hat{\v R}}{R^2} +\!\int\! d^3x'\frac{\hat{\v R}(\hat{\v R}\cdot[\partial \v J/\partial t])}{R^2c}.
\end{equation}
The integral involving $\gradv'\cdot[\v J]$ can be transformed further by examining the components of the integrand
\begin{equation}
-\frac{(\gradv'\cdot[\v J])\hat{\v R}^i}{R^2}=-\frac{\partial}{\partial x'_k}\bigg\{[\v J]_k\frac{\partial}{\partial x'_i}\bigg(\frac1R\bigg)\bigg\} + \frac{3\hat{\v R}^i(\hat{\v R} \cdot [\v J]) - [\v J]^i}{R^3}-\frac{4\pi}{3}[\v J]^i\delta(\v x-\v x'),
\end{equation}
where Eq.~(A3) has been used and summation over the index $k$ is implied. The volume integral of the first term of Eq.~(A8) can be transformed into a surface integral with the aid of Gauss' theorem, and hence vanishes assuming that the currents are localized:
\begin{equation}
\int\!d^3x'\frac{\partial}{\partial x'_k}\bigg\{[\v J]_k\frac{\partial}{\partial x'_i}\bigg(\frac1R\bigg)\bigg\}=\!\oint\!d\v S'\cdot\bigg\{[\v J]\frac{\partial}{\partial x'_i}\bigg(\frac1R\bigg)\bigg\}=0.
\end{equation}
The volume integral of the last term in Eq.~(A8) gives $-(4\pi/3)\v J$. Therefore,
\begin{equation}
-\!\int\! d^3x'\frac{(\gradv'\cdot[\v J])\hat{\v R}}{R^2}= -\frac{4\pi}{3}\v J+\!\int\! d^3x'\!\bigg(\frac{3\hat{\v R}(\hat{\v R}\!\cdot\![\v J])\!-\![\v J]}{R^3}\bigg).
\end{equation}
From Eqs.~(A7) and (A10) we obtain Eq.~(14).

\begin{acknowledgments}
The author thanks David J.\ Griffiths and Mark A.\ Heald for useful comments. The support of FICSAC is gratefully acknowledged.
\end{acknowledgments}


\begin{thebibliography}{8}
\bibitem{2} The non-self-referential form of Eq.~(2) (obtained by substituting Eq.~(3) into Eq.~(2)) implies complicated integrations (it is defined as an integral over all space, the integrand of which is itself a series of integrals over all space) and therefore it is not generally useful in practice.
\end{thebibliography}
\end{document}